\documentclass[conference]{IEEEtran}
\IEEEoverridecommandlockouts
\usepackage{cite}
\usepackage{amsmath,amssymb,amsfonts}
\usepackage{graphicx}
\usepackage{makecell}
\usepackage{textcomp}
\usepackage{xcolor}
\usepackage{adjustbox}
\usepackage{array}
\usepackage{multirow}
\usepackage{subcaption}
\usepackage{caption}
\usepackage{fancyhdr}
\fancypagestyle{IEEEtitlepagestyle}{%
  \fancyhf{}%
  \fancyhead[L]{\small X. Liu, H. Nikbakht, M. Ying, D. Shakya, H. Poddar, A. Ishii, D. Abraham, and T. S. Rappaport, ``Omnidirectional UMi Path Loss Models for 3GPP Extension Above 100 GHz Using Extensive Empirical Data Sets from 6.75 to 142 GHz,'' in \textit{2026 IEEE Military Communications Conference (MILCOM)}, Washington, DC, USA, Oct. 2026, pp. 1--6.}%
}
\newcolumntype{R}[1]{>{\raggedleft\arraybackslash}p{#1}}
\newcolumntype{C}[1]{>{\centering\arraybackslash}p{#1}}
\def\BibTeX{{\rm B\kern-.05em{\sc i\kern-.025em b}\kern-.08em
    T\kern-.1667em\lower.7ex\hbox{E}\kern-.125emX}}
\begin{document}
\title{{Omnidirectional UMi Path Loss Models for 3GPP Extension Above 100 GHz Using Extensive Empirical Data Sets from 6.75 to 142 GHz} \thanks{This paper was supported by the NYU WIRELESS Industrial Affiliates Program, NYU Tandon ECE PhD Fellowship, and NSF Grant No. 2234123.}}
\author{
\IEEEauthorblockN{
Xingchen Liu$^{1}$\IEEEauthorrefmark{1},
Homa Nikbakht$^{1}$,
Mingjun Ying$^{1}$,
Dipankar Shakya$^{1}$,
Hitesh Poddar$^{2}$,\\
Art Ishii$^{2}$,
Daniel Abraham$^{1}$,
and Theodore S. Rappaport$^{1}$\IEEEauthorrefmark{2}}
\IEEEauthorblockA{$^{1}$NYU WIRELESS, New York University, Brooklyn, NY, USA}
\IEEEauthorblockA{$^{2}$Sharp Laboratories of America, Vancouver, WA, USA}
\IEEEauthorblockA{\{xl5933\IEEEauthorrefmark{1}, tsr\IEEEauthorrefmark{2}\}@nyu.edu}
}
\bstctlcite{BSTcontrol}

\maketitle

\begin{abstract}
Extending standardized urban microcell (UMi) path loss (PL) models from the upper mid-band (FR3) to sub-terahertz (sub-THz) frequencies requires measurement-based comparisons spanning a wide spectral range within a unified framework. In this paper, we derive single- and multi-frequency omnidirectional PL models for UMi environments based on extensive NYU WIRELESS measurements at 6.75, 16.95, 28, 73, and 142~GHz in line-of-sight (LOS) and non-line-of-sight (NLOS) conditions. Single-frequency close-in (CI) and floating-intercept (FI) models for each band show that CI yields more stable and physically interpretable parameters, with shadow fading standard deviations within 0.98~dB of FI across all five frequencies. We then extend the multi-frequency PL analysis to the 7--24~GHz, 0.5--100~GHz, and 0.5--150~GHz frequency ranges using CI, close-in with frequency weighting (CIF), and alpha-beta-gamma (ABG) models. Across progressively wider frequency spans, CI and CIF maintain stable distance exponents that remain closely aligned with the 3rd Generation Partnership Project (3GPP) UMi interpretations, whereas ABG offers only very modest reductions in fitting error at the cost of much greater parameter sensitivity. These results support physically anchored CI/CIF formulations with a close-in free space path loss anchor point at 1~m in order to extend 3GPP-oriented UMi PL models over the entire 0.5 to 150~GHz frequency range.

\end{abstract}

\section{Introduction}
The demand for higher wireless capacity has driven interest in spectrum beyond the conventional sub-6 GHz global cellular bands, where much wider bandwidths are available \cite{WRC23FinalActs, rappaport2019above100ghz, Thomas2025THzSurvey}. The upper mid-band (FR3, 7--24 GHz), millimeter-wave (mmWave), and sub-terahertz (sub-THz) ranges have drawn significant research and standardization attention for 5G-Advanced (5G-A) and 6G, balancing coverage, capacity, and spectrum availability \cite{Shakya2025UMi675_1695,xing2021umi,shakya2024radio,rappaport2017overview}. Accurate urban microcell (UMi) path loss models across these bands are essential not only for commercial 6G standardization but also for public-safety communications in dense urban environments, where tactical networks increasingly rely on rapidly deployable small cells, vehicle-mounted relays, and high-capacity backhaul links operating from FR3 to sub-THz frequencies.

\begin{table*}[!t]
\vspace{0.14in}
\centering
\small
\caption{Single-frequency omnidirectional CI and FI path loss model parameters at 6.75, 16.95, 28, 73, and 142 GHz for the UMi scenario (V-V polarization) in LOS and NLOS. 95\% confidence intervals are shown in parentheses. $\Delta n = n - n_{\text{3GPP}}$, where $n_{\text{3GPP}} = 2.1$ for LOS and $n_{\text{3GPP}} = 3.19$ for NLOS using the 3GPP UMi reference \cite{TR38901}.}
\label{tab:single_freq_ci_fi_umi}
\setlength{\tabcolsep}{4.2pt}
\renewcommand{\arraystretch}{1.35}
\setcellgapes{2pt}
\makegapedcells

\begin{minipage}[t]{0.49\textwidth}
\centering
\begin{adjustbox}{max width=\linewidth,center}
\begin{tabular}{!{\vrule width 0.9pt}c|c!{\vrule width 0.7pt}c|c|c|c|c!{\vrule width 0.9pt}}
\Xhline{0.9pt}
\multicolumn{7}{!{\vrule width 0.9pt}c!{\vrule width 0.9pt}}{\textbf{Omnidirectional-LOS}} \\
\Xhline{0.9pt}
\textbf{Model}
& \textbf{Parameter}
& \makecell{\textbf{6.75}\\\textbf{GHz}}
& \makecell{\textbf{16.95}\\\textbf{GHz}}
& \makecell{\textbf{28}\\\textbf{GHz}}
& \makecell{\textbf{73}\\\textbf{GHz}}
& \makecell{\textbf{142}\\\textbf{GHz}} \\
\Xhline{0.9pt}

\multirow{3}{*}{\textbf{CI}}
& \makecell{$n$\\{\footnotesize$(\pm\,95\%\,\mathrm{conf.})$}}
& \makecell{$1.80$\\{\footnotesize$(\pm0.13)$}}
& \makecell{$1.85$\\{\footnotesize$(\pm0.20)$}}
& \makecell{$2.10$\\{\footnotesize$(\pm0.30)$}}
& \makecell{$2.01$\\{\footnotesize$(\pm0.24)$}}
& \makecell{$1.96$\\{\footnotesize$(\pm0.08)$}} \\
\cline{2-7}
& $\Delta n$
& $-0.30$
& $-0.25$
& $\phantom{+}0.00$
& $-0.09$
& $-0.14$ \\
\cline{2-7}
& $\sigma$ (dB)
& 2.57
& 4.05
& 3.52
& 4.85
& 2.65 \\
\Xhline{0.7pt}

\multirow{3}{*}{\textbf{FI}}
& $\alpha$ (dB)
& 42.71
& 41.15
& 31.85
& 115.44
& 70.11 \\
\cline{2-7}
& \makecell{$\beta$\\{\footnotesize$(\pm\,95\%\,\mathrm{conf.})$}}
& \makecell{$2.10$\\{\footnotesize$(\pm0.81)$}}
& \makecell{$2.62$\\{\footnotesize$(\pm1.10)$}}
& \makecell{$3.89$\\{\footnotesize$(\pm4.92)$}}
& \makecell{$-0.78$\\{\footnotesize$(\pm3.33)$}}
& \makecell{$2.26$\\{\footnotesize$(\pm0.74)$}} \\
\cline{2-7}
& $\sigma$ (dB)
& 2.35
& 3.13
& 2.92
& 3.87
& 2.57 \\
\Xhline{0.9pt}
\end{tabular}
\end{adjustbox}
\end{minipage}
\hfill
\begin{minipage}[t]{0.49\textwidth}
\centering
\begin{adjustbox}{max width=\linewidth,center}
\begin{tabular}{!{\vrule width 0.9pt}c|c!{\vrule width 0.7pt}c|c|c|c|c!{\vrule width 0.9pt}}
\Xhline{0.9pt}
\multicolumn{7}{!{\vrule width 0.9pt}c!{\vrule width 0.9pt}}{\textbf{Omnidirectional-NLOS}} \\
\Xhline{0.9pt}
\textbf{Model}
& \textbf{Parameter}
& \makecell{\textbf{6.75}\\\textbf{GHz}}
& \makecell{\textbf{16.95}\\\textbf{GHz}}
& \makecell{\textbf{28}\\\textbf{GHz}}
& \makecell{\textbf{73}\\\textbf{GHz}}
& \makecell{\textbf{142}\\\textbf{GHz}} \\
\Xhline{0.9pt}

\multirow{3}{*}{\textbf{CI}}
& \makecell{$n$\\{\footnotesize$(\pm\,95\%\,\mathrm{conf.})$}}
& \makecell{$2.57$\\{\footnotesize$(\pm0.21)$}}
& \makecell{$2.61$\\{\footnotesize$(\pm0.28)$}}
& \makecell{$3.42$\\{\footnotesize$(\pm0.23)$}}
& \makecell{$3.43$\\{\footnotesize$(\pm0.11)$}}
& \makecell{$2.92$\\{\footnotesize$(\pm0.33)$}} \\
\cline{2-7}
& $\Delta n$
& $-0.62$
& $-0.58$
& $+0.23$
& $+0.24$
& $-0.27$ \\
\cline{2-7}
& $\sigma$ (dB)
& 6.53
& 8.72
& 9.75
& 7.87
& 8.30 \\
\Xhline{0.7pt}

\multirow{3}{*}{\textbf{FI}}
& $\alpha$ (dB)
& 51.69
& 68.12
& 80.62
& 80.57
& 120.57 \\
\cline{2-7}
& \makecell{$\beta$\\{\footnotesize$(\pm\,95\%\,\mathrm{conf.})$}}
& \makecell{$2.44$\\{\footnotesize$(\pm1.29)$}}
& \makecell{$2.11$\\{\footnotesize$(\pm1.68)$}}
& \makecell{$2.50$\\{\footnotesize$(\pm3.50)$}}
& \makecell{$2.89$\\{\footnotesize$(\pm1.33)$}}
& \makecell{$0.36$\\{\footnotesize$(\pm4.29)$}} \\
\cline{2-7}
& $\sigma$ (dB)
& 6.51
& 8.50
& 9.66
& 7.82
& 7.56 \\
\Xhline{0.9pt}
\end{tabular}
\end{adjustbox}
\end{minipage}
\end{table*}

Standardization for these emerging bands is progressing, but important gaps remain. The 3rd Generation Partnership Project (3GPP) Technical Report (TR)~38.901 \cite{TR38901} defines channel models from 0.5 to 100 GHz for key scenarios, including the UMi street canyon, and has been the 5G baseline since Release 14. However, it does not cover frequencies above 100 GHz, and no standardized UMi path loss reference yet exists for the sub-THz range. WRC-23 recently identified candidate IMT-2030 spectrum, including sub-THz bands \cite{WRC23FinalActs}, and both 3GPP \cite{Poddar2026Measurement,Poddar2025Overview} and the ITU are extending their channel modeling toward 6G. As models extend from FR3 through mmWave to sub-THz, the goal is path loss expressions with stable, physically interpretable parameters compatible with the existing UMi framework.

In this paper, we address these gaps by analyzing omnidirectional UMi path loss measurements collected by NYU WIRELESS at 6.75, 16.95, 28, 73, and 142 GHz at NYU urban sites in Manhattan and Brooklyn from 2012 to 2024 \cite{Azar2013UMi28, itwillwork, maccartney2019millimeter, MacCartney2014UMi73, xing2021millimeter, Shakya2025UMi675_1695}. The unified analysis applies a common omnidirectional co-polarized vertical-to-vertical (V-V) representation separately to the line-of-sight (LOS) and non-line-of-sight (NLOS) data from all five campaigns, enabling a nested-span test of parameter stability as measured bands are added. While each campaign has been reported separately \cite{Azar2013UMi28, itwillwork, maccartney2019millimeter, MacCartney2014UMi73, xing2021millimeter, Shakya2025UMi675_1695}, this paper combines the datasets to provide a unified multi-band UMi analysis spanning FR3, mmWave, and sub-THz, with TX-RX distances ranging from 24 to 880~m. From these data, we derive single- and multi-frequency PL parameters, report 95\% confidence intervals on the PL slope parameters for statistical comparison with 3GPP, and evaluate model stability as the fitted frequency spans from FR3 to 142 GHz.
The main contributions of this paper are:
\begin{itemize}
    \item Single-frequency close-in (CI) and floating-intercept (FI) PL parameters with 95\% confidence intervals are derived at 6.75, 16.95, 28, 73, and 142 GHz for LOS and NLOS conditions. The 142 GHz results directly test 3GPP-oriented UMi modeling beyond the 0.5--100 GHz scope of TR~38.901~\cite{TR38901}.
    \item Using the massive empirical datasets, multi-frequency CI, close-in with frequency weighting (CIF), and alpha-beta-gamma (ABG) parameters with 95\% confidence intervals are derived over three frequency ranges: 7--24 GHz, 0.5--100 GHz, and 0.5--150 GHz.
    \item The CI/CIF NLOS exponents over the wider frequency ranges are shown to be statistically consistent with the 3GPP UMi-NLOS reference ($n=3.19$), whereas the ABG frequency exponent is noisier and more sensitive to the measurement frequency sets.
\end{itemize}

A surprising result is that the extensive empirical data support CI/CIF PL models as a more stable basis for UMi PL modeling consistent with the existing 3GPP framework above 100 GHz.
The remainder of this paper is organized as follows. Section~II describes the measurement campaigns. Sections~III and IV present the single-frequency and multi-frequency PL models, respectively. Section~V concludes the paper.

\section{Overview of Measurements}

Omnidirectional UMi PL measurements in LOS and NLOS were collected by NYU WIRELESS at 6.75, 16.95, 28, 73, and 142 GHz using sliding-correlation channel sounders (1 GHz bandwidth, except 800~MHz at 28 GHz) \cite{Azar2013UMi28, MacCartney2014UMi73, Shakya2025UMi675_1695, xing2021umi, itwillwork, rappaport2015wideband, rappaport2019above100ghz, ju2021spatial142ghz}. The 28 GHz (2012, 46.9~GB) \cite{Azar2013UMi28, itwillwork} and 73 GHz (2013, 34.4~GB) \cite{MacCartney2014UMi73, rappaport2015wideband} campaigns around NYU's Manhattan campus yielded 25 TX-RX pairs (5 LOS, 20 NLOS) over 31--186~m and 62 pairs (9 LOS, 53 NLOS) over 27--190~m, respectively, with TX heights of 7/17~m and RX heights of 1.5--4.06~m. The 142 GHz campaign (2020, 85.2~GB) \cite{xing2021umi, ju2021spatial142ghz, rappaport2019above100ghz} at MetroTech Commons, Brooklyn yielded 27 pairs (16 LOS, 11 NLOS) over 24--117~m, and the 6.75 and 16.95 GHz campaigns (2024, 12.9 and 24.3~GB) \cite{Shakya2025UMi675_1695} at the same site each yielded 18 pairs (7 LOS, 11 NLOS) over 40--880~m, all with TX/RX heights of 4~m/1.5~m. Together the five campaigns, comprising over 200~GB of raw wideband data, provide a comprehensive measurement basis for unified UMi PL modeling from FR3 to sub-THz~\cite{shakya2025milcom}.

\section{Single-Frequency Path Loss Models}

Single-frequency PL models characterize large-scale signal attenuation over distance at a given carrier frequency. Here, two standard single-frequency formulations are considered, namely the CI free-space reference distance model and the FI model \cite{sun2016investigation, rappaport2017overview}.

The CI PL model anchors large-scale signal attenuation to a close-in free space path loss (FSPL) reference distance $d_0$ (which is ideally, but not necessarily, in the far field) \cite{rappaport2002wireless, rappaport2015mmwavebook, sun2016investigation, rappaport2017overview}, and describes the excess attenuation beyond free space through a single parameter, the path loss exponent (PLE) $n$. The CI model assumes $d_0=1$~m and is expressed as \cite{rappaport2002wireless, rappaport2015mmwavebook}
\begin{equation}
\mathrm{PL}^{\mathrm{CI}}(f,d) [\mathrm{dB}]=\mathrm{FSPL}(f,d_0)+10n\log_{10}\!\left(\frac{d}{d_0}\right)+X_{\sigma}^{\mathrm{CI}},
\label{eq:ci_single_umi}
\end{equation}
where
\begin{equation}
\mathrm{FSPL}(f,d_0)[\mathrm{dB}]=32.4+20\log_{10}(f),
\label{eq:fspl_single_umi}
\end{equation}
for $f$ in GHz and $d_0=1$ m, where $d$ is the TX-RX separation distance in meters, $n$ is the PLE, and $X_{\sigma}^{\mathrm{CI}}\sim\mathcal{N}(0,\sigma^2)$ models shadow fading in dB; the 32.4~dB constant is the free-space path loss at 1~m and 1 GHz \cite{rappaport2002wireless, rappaport2015mmwavebook}. Due to the explicit FSPL anchor at $d_0=1$~m, the CI model separates the frequency dependence in the first-meter free-space term from the distance dependence captured by the PLE.

In contrast, the FI PL model does not impose a physical reference point, but instead fits both the intercept and slope directly from the measured data, such that there is no anchor for the curve-fitting at a close-in reference. The 3GPP community has generally adopted the FI model \cite{sun2016investigation, TR38901}, as
\begin{equation}
\mathrm{PL}^{\mathrm{FI}}(d)[\mathrm{dB}]=\alpha+10\beta\log_{10}(d)+X_{\sigma}^{\mathrm{FI}},
\label{eq:fi_single_umi}
\end{equation}
where $\alpha$ is the floating intercept in dB (which has no basis in physics and varies widely with the measurement set, as Table~\ref{tab:single_freq_ci_fi_umi} demonstrates), $\beta$ is the fitted distance-dependent PL slope (which would ideally have a value similar to the CI PLE $n$), and $X_{\sigma}^{\mathrm{FI}}\sim\mathcal{N}(0,\sigma^2)$ denotes shadow fading in dB \cite{sun2016vtc_pl}.
 All parameters are obtained via minimum mean square error linear regression, and the 95\% confidence intervals reported throughout are computed from the regression standard errors using the Student's $t$ distribution \cite{montgomery2012linear}.

The 3GPP TR~38.901 UMi Street Canyon models \cite{TR38901} provide reference exponents for interpreting the fitted CI parameters. The UMi CI model covers 0.5 to 100 GHz and uses $d_0=1$~m in~\eqref{eq:ci_single_umi}, with $n=2.1$ and $\sigma=4.0$~dB for LOS. For NLOS, the 3GPP UMi CI reference is $n=3.19$ with $\sigma=8.2$~dB. Accordingly, the reported $\Delta n = n - n_{\text{3GPP}}$ values, i.e., the difference between the measured CI PLE and the 3GPP reference exponent, are computed relative to $n_{\text{3GPP}}=2.1$ for LOS and $n_{\text{3GPP}}=3.19$ for NLOS \cite{TR38901}.

Table~\ref{tab:single_freq_ci_fi_umi} summarizes the CI (with $d_0=1$~m) and FI parameters fitted at the five frequencies using vertical-vertical, co-polarized (V-V) measurement data. Over the measured links, the CI and FI residual shadow fading standard deviations differ by at most 0.98~dB at every frequency. The CI PLE varies in a narrow and physically reasonable range (1.80 to 2.10 in LOS and 2.57 to 3.43 in NLOS), tracking the 3GPP references closely with $|\Delta n|\leq 0.30$ in LOS and $|\Delta n|\leq 0.62$ in NLOS. The FI parameters, by contrast, fluctuate dramatically: the intercept $\alpha$ swings from 31.85 to 115.44~dB in LOS and 51.69 to 120.57~dB in NLOS, while the slope $\beta$ ranges from 3.89 down to $-0.78$. The 95\% confidence intervals on $\beta$ are substantially wider than those of $n$ and often span zero or change sign, e.g., $\beta = 3.89 \pm 4.92$ at 28 GHz LOS, $\beta = -0.78 \pm 3.33$ at 73 GHz LOS, and $\beta = 0.36 \pm 4.29$ at 142 GHz NLOS. Thus, CI achieves comparable in-sample residual dispersion with substantially more stable and physically interpretable parameters across the five measured bands.
\begin{figure}[t]
    \vspace{0.03in}
    \centering
    \includegraphics[width=0.9\linewidth]{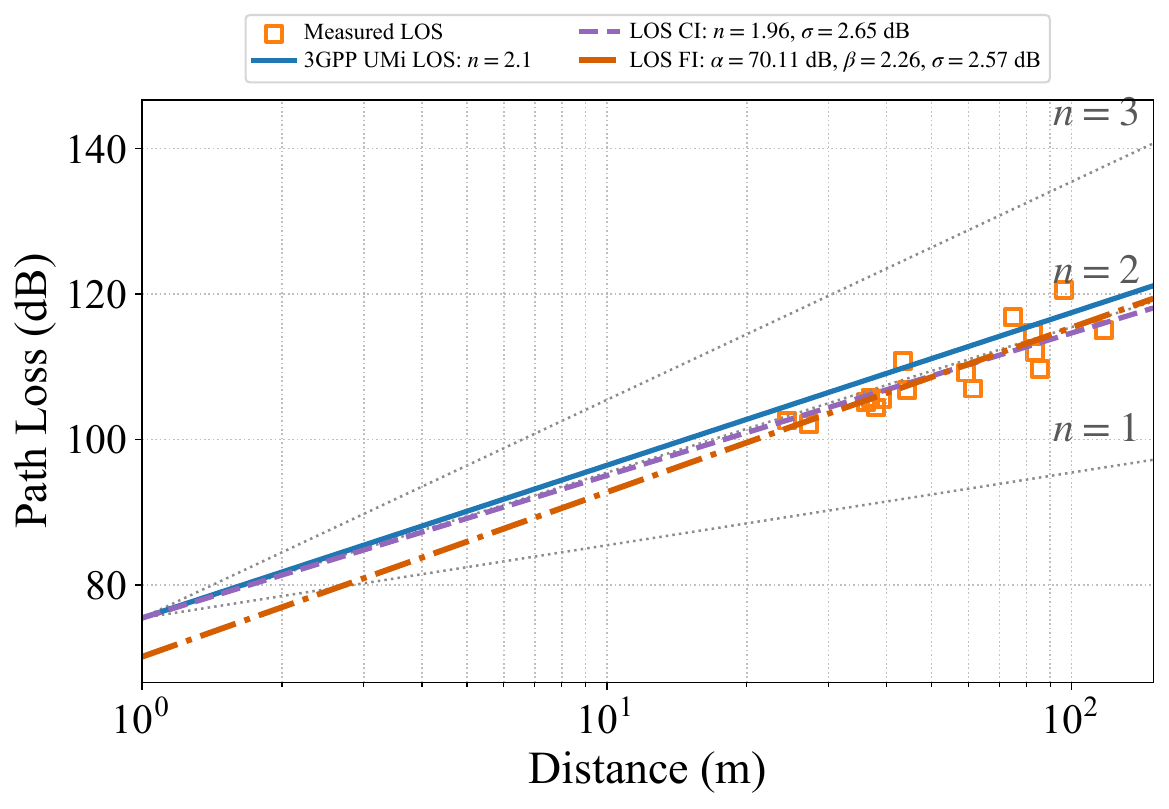}
\caption{Measured 142 GHz UMi LOS path loss with CI and FI fits and the 3GPP UMi-LOS reference ($n=2.1$) extrapolated beyond the 0.5--100 GHz scope of TR~38.901 \cite{TR38901}. The CI and FI residual standard deviations are 2.65 and 2.57~dB, respectively.}
    \label{fig:umi_142_los_singlefreq}
\end{figure}
\begin{figure}[!t]
    \vspace{0.03in}
    \centering
    \includegraphics[width=0.9\linewidth]{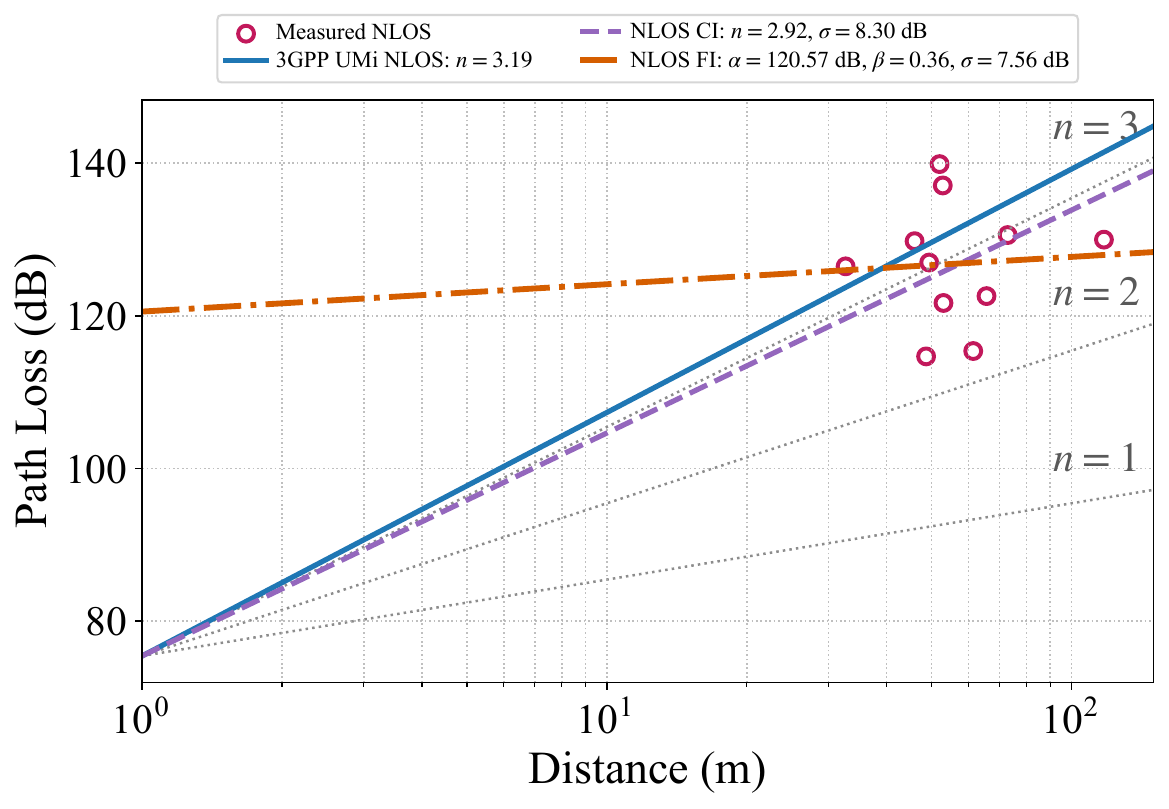}
    \caption{Measured 142 GHz UMi NLOS path loss with CI and FI fits and the optional 3GPP UMi-NLOS reference ($n=3.19$) extrapolated beyond the 0.5--100 GHz scope of TR~38.901 \cite{TR38901}. The CI fit retains $n=2.92$, whereas the FI fit is nearly flat ($\beta=0.36$) because the 120.57~dB FI intercept absorbs most of the distance trend.}
    \label{fig:umi_142_nlos_singlefreq}
\end{figure}
These FI instabilities reflect the known behavior of floating-intercept regression \cite{sun2016investigation}: the intercept absorbs part of the free-space loss and distance trend, leaving the slope sensitive to measurement geometry, an effect amplified with few locations (e.g., the negative $\beta$ at $73$ GHz LOS is a 9-pair fitting artifact). Since the 1~m FSPL increases by only 26.5~dB across 6.75--142 GHz while $\alpha$ swings by over 80~dB, CI is the more reliable single-frequency formulation.

The contrast between the CI and FI fits is particularly clear at 142 GHz. Figs.~\ref{fig:umi_142_los_singlefreq} and~\ref{fig:umi_142_nlos_singlefreq} include 3GPP-based UMi reference curves obtained by evaluating the TR~38.901 LOS and NLOS formulations at $f_c = 142$ GHz \cite{TR38901}, used as extrapolated references only. In LOS, both the CI and FI fits deviate by less than 0.16 in distance dependence from the reference. The difference is more evident in NLOS: the CI fit ($n = 2.92$) and the 3GPP NLOS reference ($n = 3.19$) both retain clear distance dependence and agree closely, while the FI fit becomes nearly flat ($\beta = 0.36$). Without an FSPL anchor, $\alpha = 120.57$~dB absorbs most of the propagation loss, leaving $\beta$ with insufficient residual distance dependence for realistic outdoor conditions. The residual standard deviations characterize the measured 142 GHz distance range of 24--117~m. Below 24~m, the plotted curves are extrapolations that expose the intercept--slope sensitivity of FI. Overall, the single-frequency CI model preserves a clear physical anchor through the FSPL at $1$~m and yields PLEs consistent with 3GPP UMi references from FR3 to sub-THz, motivating the multi-frequency analysis next.
Prior NYU WIRELESS propagation research spans indoor coverage computation, building-database and ray-tracing methods, mmWave indoor/backhaul measurements and multi-beam combining, and multi-frequency outdoor statistical channel modeling~\cite{panjwani1996coverage,rappaport2004buildingDatabase,rappaport2004receptionSurfaces,nie2013indoor,sun2014multibeam,samimi2015multifreq}.

\begin{table*}[t]
\vspace{0.14in}
\centering
\small
\caption{Multi-frequency omnidirectional CI, CIF, and ABG parameters for UMi V-V measurements over 7--24 GHz, 0.5--100 GHz, and 0.5--150 GHz, alongside 3GPP TR~38.901 UMi reference values \cite{TR38901}. Parentheses denote 95\% confidence intervals; $\Delta$ terms denote differences from the corresponding 3GPP reference; and ``--'' indicates an undefined or inapplicable parameter.}
\label{tab:mf_combined}
\setlength{\tabcolsep}{6pt}
\renewcommand{\arraystretch}{1.25}
\begin{adjustbox}{max width=\textwidth,center}
\begin{tabular}{!{\vrule width 1.0pt}c|c!{\vrule width 0.8pt}c|c|c|c!{{\vrule width 0.7pt}\hskip 1.5pt{\vrule width 0.7pt}}c|c|c|c!{\vrule width 1.0pt}}
\Xhline{1.0pt}
\multirow{2}{*}{\textbf{Model}}
& \multirow{2}{*}{\textbf{Parameter}}
& \multicolumn{4}{c!{{\vrule width 0.7pt}\hskip 1.5pt{\vrule width 0.7pt}}}{\textbf{Omnidirectional-LOS}}
& \multicolumn{4}{c!{\vrule width 1.0pt}}{\textbf{Omnidirectional-NLOS}} \\
\cline{3-10}
\rule{0pt}{3.0ex}&
& \makecell{\textbf{7--24}\\\textbf{GHz}}
& \makecell{\textbf{0.5--100}\\\textbf{GHz}}
& \makecell{\textbf{0.5--150}\\\textbf{GHz}}
& \makecell{\textbf{3GPP}~\cite{TR38901}\\\textbf{0.5--100}}
& \makecell{\textbf{7--24}\\\textbf{GHz}}
& \makecell{\textbf{0.5--100}\\\textbf{GHz}}
& \makecell{\textbf{0.5--150}\\\textbf{GHz}}
& \makecell{\textbf{3GPP}~\cite{TR38901}\\\textbf{0.5--100}} \\
\Xhline{1.0pt}

\multirow{3}{*}{\textbf{CI}}
& $n${\scriptsize$\,(\pm95\%)$}
& $1.82${\scriptsize$\,(\pm0.10)$} & $1.91${\scriptsize$\,(\pm0.10)$} & $1.93${\scriptsize$\,(\pm0.07)$} & $2.1$
& $2.59${\scriptsize$\,(\pm0.16)$} & $3.21${\scriptsize$\,(\pm0.11)$} & $3.19${\scriptsize$\,(\pm0.10)$} & $3.19$ \\
\cline{2-10}
& $\Delta n$
& $-0.28$ & $-0.19$ & $-0.17$ & --
& $-0.60$ & $+0.02$ & $\phantom{+}0.00$ & -- \\
\cline{2-10}
& $\sigma$ (dB)
& 3.43 & 4.47 & 3.93 & 4.0
& 7.72 & 11.21 & 11.05 & 8.2 \\
\Xhline{1.0pt}

\multirow{4}{*}{\textbf{CIF}}
& $n${\scriptsize$\,(\pm95\%)$}
& $1.82${\scriptsize$\,(\pm0.10)$} & $1.92${\scriptsize$\,(\pm0.09)$} & $1.93${\scriptsize$\,(\pm0.07)$} & --
& $2.59${\scriptsize$\,(\pm0.16)$} & $3.22${\scriptsize$\,(\pm0.10)$} & $3.20${\scriptsize$\,(\pm0.10)$} & -- \\
\cline{2-10}
& $b$
& 0.03 & 0.06 & 0.03 & --
& 0.02 & 0.16 & 0.07 & -- \\
\cline{2-10}
& $f_0$ (GHz)
& 12 & 34 & 74 & --
& 12 & 49 & 59 & -- \\
\cline{2-10}
& $\sigma$ (dB)
& 3.39 & 4.21 & 3.84 & --
& 7.70 & 9.57 & 10.66 & -- \\
\Xhline{1.0pt}

\multirow{7}{*}{\textbf{ABG}}
& $\alpha${\scriptsize$\,(\pm95\%)$}
& $2.35${\scriptsize$\,(\pm0.58)$} & $2.03${\scriptsize$\,(\pm0.64)$} & $1.97${\scriptsize$\,(\pm0.47)$} & $2.10$
& $2.28${\scriptsize$\,(\pm0.96)$} & $2.41${\scriptsize$\,(\pm0.81)$} & $2.87${\scriptsize$\,(\pm0.85)$} & $3.19$ \\
\cline{2-10}
& $\beta$
& 19.36 & 23.20 & 27.36 & 32.4
& 36.15 & 22.94 & 20.55 & 32.4 \\
\cline{2-10}
& $\gamma$
& 2.20 & 2.51 & 2.26 & 2.0
& 2.31 & 3.66 & 3.14 & 2.0 \\
\cline{2-10}
& $\Delta\alpha$
& $+0.25$ & $-0.07$ & $-0.13$ & --
& $-0.91$ & $-0.78$ & $-0.32$ & -- \\
\cline{2-10}
& $\Delta\beta$
& $-13.04$ & $-9.20$ & $-5.04$ & --
& $+3.75$ & $-9.46$ & $-11.85$ & -- \\
\cline{2-10}
& $\Delta\gamma$
& $+0.20$ & $+0.51$ & $+0.26$ & --
& $+0.31$ & $+1.66$ & $+1.14$ & -- \\
\cline{2-10}
& $\sigma$ (dB)
& 2.90 & 4.06 & 3.73 & 4.0
& 7.60 & 9.04 & 10.02 & 8.2 \\
\Xhline{1.0pt}
\end{tabular}
\end{adjustbox}
\end{table*}

\section{Multi-Frequency Path Loss Models}

Multi-frequency path loss models describe large-scale attenuation as a joint function of TX--RX separation and carrier frequency. A useful wideband model should not only achieve a close fit to the measured data, but also produce parameters that remain stable and physically interpretable as the fitted frequency span is extended, especially when FR3, mmWave, and sub-THz measurements are combined in a single model.

CI, CIF, and ABG provide a controlled comparison of one-, two-, and three-parameter multi-frequency models with and without a physical FSPL anchor.
The multi-frequency CI model extends the single-frequency CI formulation by fitting a single PLE across the frequencies of interest. Based on this, the CIF model introduces an additional frequency-weighting term into the distance-dependent component of the CI model \cite{sun2016investigation, sun2016vtc_pl, rappaport2017overview}, while still retaining the FSPL anchor at reference distance. In this sense, CIF can be viewed as a physically anchored extension of CI for cases where the effective path loss exponent may vary slightly with frequency. The model is given by
\begin{equation}
\begin{split}
\mathrm{PL}^{\mathrm{CIF}}&(f,d)\,[\mathrm{dB}]
= \mathrm{FSPL}(f,d_0) \\
& + 10n\!\left[1+b\!\left(\frac{f-f_0}{f_0}\right)\right]\!\log_{10}\!\left(\frac{d}{d_0}\right) + X_{\sigma}^{\mathrm{CIF}}.
\end{split}
\label{eq:mf_cif}
\end{equation}
where $b$ captures the additional frequency weighting, $f_0$ is the weighted reference frequency, and $n$ is the distance exponent. Specifically, $f_0=\sum_k N_k f_k / \sum_k N_k$, where $N_k$ is the number of measurement points at campaign frequency $f_k$ in the fitted span~\cite{sun2016investigation}; the calculated $f_0$ is rounded to the nearest integer in GHz in this work. A positive $b$ indicates that the effective PLE increases with frequency relative to $f_0$, reflecting additional frequency-dependent attenuation beyond what the FSPL term already captures, whereas a negative $b$ would imply a decreasing effective PLE at higher frequencies. When $b=0$, the CIF model reduces to the CI model.

As a more empirical alternative, the ABG model represents path loss through separate fitted coefficients for the distance dependence, the frequency dependence, and an overall offset~\cite{TR38901}. Unlike CI and CIF, the ABG model does not preserve explicit FSPL at reference distance, but instead offers greater flexibility in jointly fitting data. The ABG model is expressed as
\begin{equation}
\begin{split}
\mathrm{PL}^{\mathrm{ABG}}(f,d)[\mathrm{dB}]
&=
10\alpha\log_{10}\!\left(\frac{d}{d_0}\right)
+\beta \\
&\quad +10\gamma\log_{10}\!\left(\frac{f}{1~\mathrm{GHz}}\right)
+X_{\sigma}^{\mathrm{ABG}},
\end{split}
\label{eq:mf_abg}
\end{equation}
where $\alpha$ and $\gamma$ characterize the distance and frequency dependence, respectively, and $\beta$ is an optimized offset.

The three model classes also differ in how directly they connect to the standardized UMi formulation. Section~III compared the models against TR~38.901 at single frequencies; in the multi-frequency setting, a model must instead match TR~38.901 across both distance and frequency simultaneously.
Below the breakpoint distance, the 3GPP UMi LOS expression \cite{TR38901} corresponds to a CI model with $n=2.1$ and to an ABG model with $(\alpha,\beta,\gamma)=(2.1,32.4,2)$. The 3GPP UMi-NLOS expression \cite{TR38901} corresponds likewise to a CI model with $n=3.19$ and to an ABG model with $(\alpha,\beta,\gamma)=(3.19,32.4,2)$.

\begin{figure*}[t]
    \centering
    \begin{subfigure}[t]{0.45\textwidth}
        \centering
        \includegraphics[width=\linewidth]{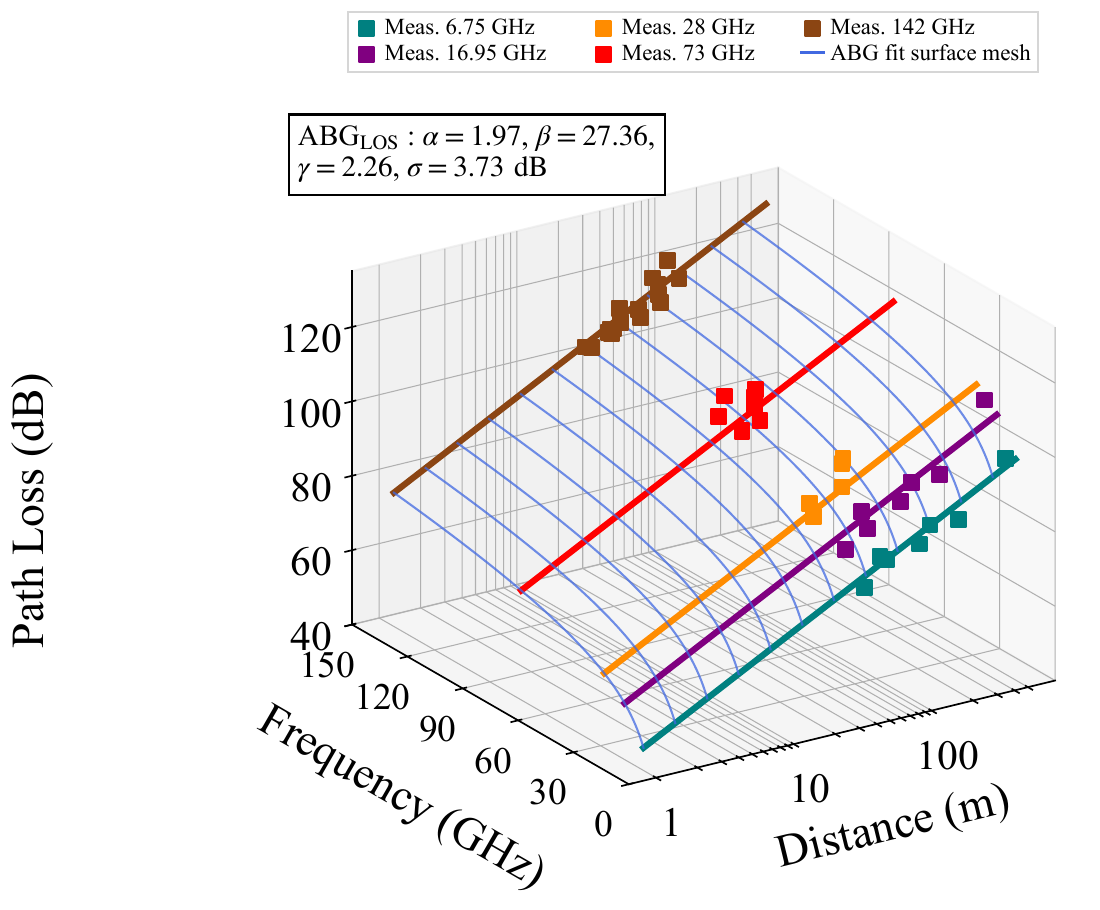}
        \caption{UMi LOS.}
        \label{fig:abg_3d_los_vv}
    \end{subfigure}
    \hfill
    \begin{subfigure}[t]{0.45\textwidth}
        \centering
        \includegraphics[width=\linewidth]{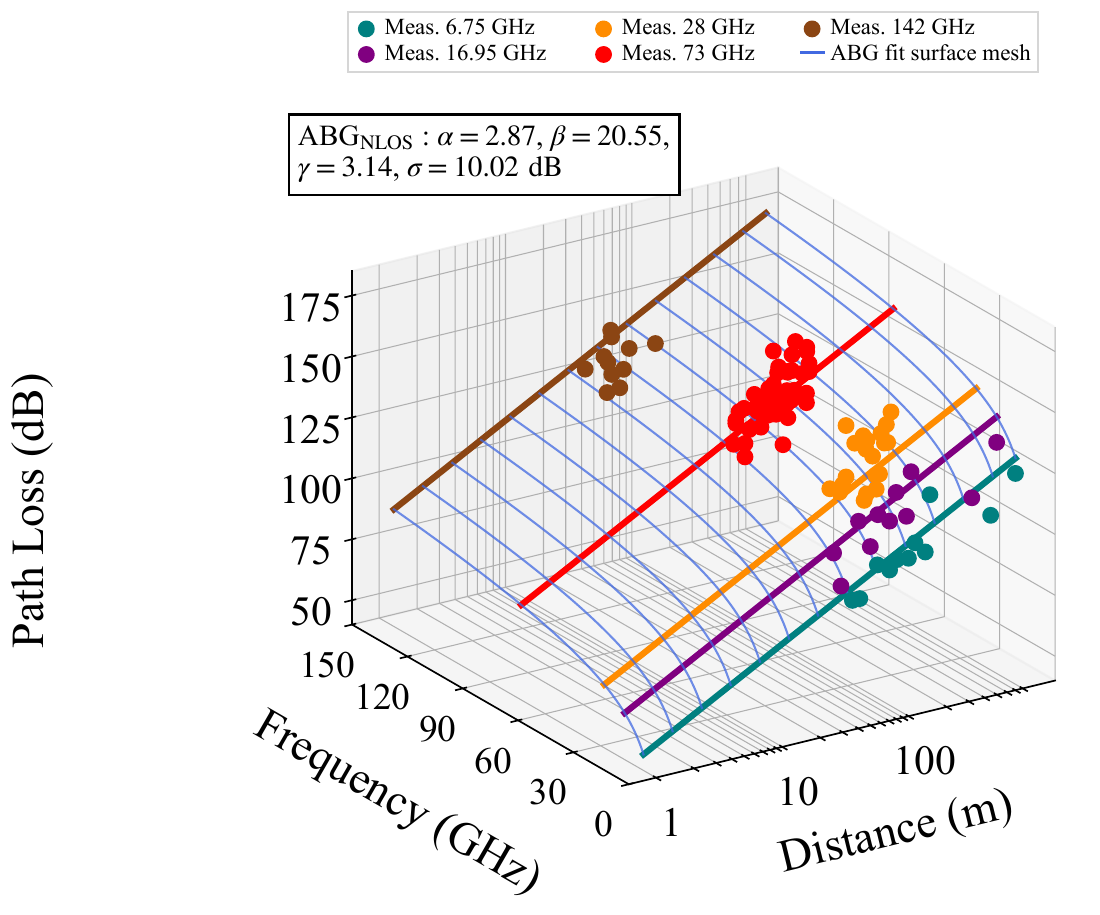}
        \caption{UMi NLOS.}
        \label{fig:abg_3d_nlos_vv}
    \end{subfigure}
  \caption{ABG fits to pooled UMi V-V measurements over 0.5--150 GHz with 3GPP reference surfaces \cite{TR38901}: (a) LOS and (b) NLOS. Markers show the five measured bands; solid cuts and the blue mesh show the fitted ABG model; dotted cuts and the translucent surface show the 3GPP references.}
    \label{fig:abg_3d_vv}
\end{figure*}

We fit the multi-frequency models over three progressively wider frequency spans. The first is 7--24 GHz, represented by the 6.75 and 16.95 GHz measurements. The second is 0.5--100 GHz, using 6.75, 16.95, 28, and 73 GHz data, which enables direct comparison over the principal range covered by TR~38.901. The third is 0.5--150 GHz, which adds the independently conducted 142 GHz campaign. The nested spans test parameter stability as additional independently measured campaigns are incorporated, including the 142 GHz band above 100 GHz.

Table~\ref{tab:mf_combined} summarizes the fitted multi-frequency parameters over the three frequency spans using V-V measurement data. Among the three models, CI uses a single fitted parameter (the PLE $n$) with an FSPL anchor; CIF extends CI with one additional frequency-weighting parameter $b$; and ABG fits three independent parameters ($\alpha$, $\beta$, $\gamma$) without a physical anchor. Despite these differences in model complexity, the CI, CIF, and ABG models yield comparable shadow fading standard deviations across all three spans. ABG generally produces the smallest $\sigma$, but the reduction relative to CIF, the most directly comparable physically anchored model, is small: 0.49, 0.15, and 0.11~dB in LOS and 0.10, 0.53, and 0.64~dB in NLOS for the 7--24, 0.5--100, and 0.5--150 GHz spans, respectively. The residual values summarize dispersion over the pooled measurement support, whereas the 95\% confidence intervals quantify parameter-estimation uncertainty. Across all three models, NLOS $\sigma$ is 1.44--3.49~dB larger for the two wider spans than for 7--24 GHz; LOS $\sigma$ remains nonmonotonic. The additional flexibility of ABG therefore yields only marginal gains in fitting accuracy.

A clearer distinction emerges in the fitted distance dependence and the corresponding $\Delta n$ values. For the CI model, the LOS PLE changes only slightly as the span widens, increasing from 1.82 over 7--24 GHz to 1.91 over 0.5--100 GHz and 1.93 over 0.5--150 GHz. Correspondingly, the magnitude of $\Delta n$ relative to the 3GPP UMi LOS reference decreases from $-0.28$ to $-0.19$ and $-0.17$, indicating that the measured LOS distance dependence remains close to, but slightly below, the standardized UMi LOS reference. The NLOS trend is more pronounced. The CI PLE increases from 2.59 in the 7--24 GHz span to 3.21 over 0.5--100 GHz and 3.19 over 0.5--150 GHz, giving $\Delta n=-0.60$, $+0.02$, and $0.00$ relative to the 3GPP UMi-NLOS reference of $n=3.19$.
The narrow 7--24 GHz fit therefore yields an NLOS exponent well below the standardized reference, whereas the wider multi-frequency fits converge almost exactly onto the 3GPP UMi-NLOS reference.

The CI/CIF interpretation is further supported by the CIF frequency-weighting parameter $b$, which remains small throughout. Over 7--24 GHz, $b$ is only 0.03 in LOS and 0.02 in NLOS.
Over 0.5--100 GHz, $b$ rises to 0.06 and 0.16, and once the 142 GHz data are included in the 0.5--150 GHz fit, $b$ falls back to 0.03 and 0.07.
Thus, once the FSPL anchor is retained, only weak additional frequency weighting is required to represent the measured UMi data.
Most of the frequency dependence is already captured by the free-space term itself, reinforcing the case for CI/CIF formulations as a basis for physically anchored wideband modeling.

The ABG model exhibits a different parameter evolution across the three frequency spans.
In LOS, $\alpha$ decreases from 2.35 to 1.97 and $\gamma$ varies between 2.20 and 2.51 across the three spans. In NLOS, $\alpha$ increases from 2.28 to 2.87 and $\gamma$ shifts from 2.31 to 3.66 and then to 3.14 as the fitted frequency range is extended to include 142 GHz. The relatively larger variation in $\gamma$ compared to $\alpha$ reflects the sparse frequency sampling inherent in each fitted range: while $\alpha$ is estimated from measurements distributed over the full TX-RX distance range within each campaign, $\gamma$ is determined by only 2, 4, or 5 distinct frequency points in the 7--24, 0.5--100, and 0.5--150 GHz fits, respectively. With so few frequency samples, the inclusion of a new frequency band can substantially shift the fitted $\gamma$, exposing a key limitation of the ABG model.

The model sensitivity across the three fitted spans further distinguishes them. The CI and CIF distance exponents vary by only 0.11 in LOS and 0.63 in NLOS, and the CIF term $b$ by at most 0.14, while $\sigma$ stays within about 1~dB (LOS) and 3.5~dB (NLOS). By contrast, although ABG attains a comparable or slightly smaller $\sigma$, its offset $\beta$ varies by 8.00~dB (LOS) and 15.60~dB (NLOS) and its exponent $\gamma$ by up to 1.35, far more sensitive to the fitted frequency set than CI/CIF. Of the two physically anchored models, CIF is preferable: it retains CI's stable distance exponent yet yields a lower $\sigma$ in every span, varying by only 0.82~dB in LOS and 2.96~dB in NLOS, versus 1.04~dB and 3.49~dB for CI.

Fig.~\ref{fig:abg_3d_vv} provides a complementary visualization of the pooled V-V measurements over 0.5--150 GHz using the ABG model, with the 3GPP UMi reference surface included for comparison. In LOS, the fitted ABG mesh and the 3GPP UMi reference surface are nearly coincident. In NLOS, the two surfaces show visible separation, with the fitted ABG surface lying above the 3GPP reference, consistent with the parameter shifts discussed above.

Taken together, the wider-span CI/CIF NLOS exponents are statistically consistent with the 3GPP UMi-NLOS exponent of $n=3.19$, which falls within the 95\% confidence intervals of the measured estimates. Adding the 142 GHz campaign directly tests parameter stability when the measured range extends beyond 100 GHz; ABG remains the residual-error reference over the same spans.

\section{Conclusion}
\label{sec:conclusion}
We presented single- and multi-frequency omnidirectional UMi path loss models based on extensive NYU WIRELESS wide-band measurements at 6.75, 16.95, 28, 73, and 142 GHz. Adding the independently conducted 142 GHz campaign changes the CI and CIF distance exponents by at most 0.02 in both LOS and NLOS. Over the same extension, ABG lowers $\sigma$ by only 0.11~dB in LOS and 0.64~dB in NLOS while shifting $\alpha$ by 0.46 in NLOS and $\beta$ by 2.39--4.16~dB. The five-band evidence therefore supports physically anchored CI/CIF formulations as candidates for 3GPP-oriented UMi modeling at 142 GHz. Future work includes directional modeling, atmospheric absorption above 100 GHz, and validation with additional campaigns~\cite{ying2025locationOptimization,ying2026nyurayCalibration}, including ongoing NYU WIRELESS measurements at 3.7, 180, 220, and 280 GHz.

\bibliographystyle{IEEEtran}
\bibliography{refs}

@IEEEtranBSTCTL{BSTcontrol,
  CTLuse_forced_etal = "yes",
  CTLmax_names_forced_etal = 1,
  CTLnames_show_etal = 1,
  CTLdash_repeated_names = "no"
}

@inproceedings{Shakya2025UMi675_1695,
  author    = {Dipankar Shakya and Mingjun Ying and Theodore S. Rappaport and Peijie Ma and Idris Al-Wazani and Yanze Wu and Yanbo Wang and Doru Calin and Hitesh Poddar and Ahmad Bazzi and Marwa Chafii and Yunchou Xing and Amitava Ghosh},
  title     = {Urban Outdoor Propagation Measurements and Channel Models at 6.75 {GHz FR1(C)} and 16.95 {GHz FR3} Upper Mid-Band Spectrum for {5G} and {6G}},
  booktitle = {Proc. IEEE Int. Conf. Commun. (ICC)},
  year      = {2025},
  pages     = {3291--3296},
  doi       = {10.1109/ICC52391.2025.11161884}
}

@inproceedings{Azar2013UMi28,
  author    = {Yaniv Azar and George N. Wong and Kevin Wang and Rimma Mayzus and Jocelyn K. Schulz and Hang Zhao and Felix {Gutierrez Jr.} and DuckDong Hwang and Theodore S. Rappaport},
  title     = {28 {GHz} Propagation Measurements for Outdoor Cellular Communications Using Steerable Beam Antennas in {New York City}},
  booktitle = {Proc. IEEE Int. Conf. Commun. (ICC)},
  year      = {2013},
  pages     = {5143--5147},
  doi       = {10.1109/ICC.2013.6655399}
}

@inproceedings{MacCartney2014UMi73,
  author    = {George R. {MacCartney Jr.} and Theodore S. Rappaport},
  title     = {73 {GHz} Millimeter Wave Propagation Measurements for Outdoor Urban Mobile and Backhaul Communications in {New York City}},
  booktitle = {Proc. IEEE Int. Conf. Commun. (ICC)},
  year      = {2014},
  pages     = {4862--4867},
  doi       = {10.1109/ICC.2014.6884090}
}

@misc{WRC23FinalActs,
  author       = {{International Telecommunication Union}},
  title        = {Final Acts of the World Radiocommunication Conference {(WRC-23)}},
  howpublished = {ITU-R Official Document},
  year         = {2023},
}

@misc{Poddar2025Overview,
  author       = {H. Poddar and D. Gold and D. Lee and N. Zhang and
                  G. Sridharan and H. Asplund and M. Shafi},
  title        = {Overview of {3GPP} Release 19 Study on Channel Modeling
                  Enhancements to {TR~38.901} for {6G}},
  howpublished = {arXiv:2507.19266},
  year         = {2025},
}

@misc{Poddar2026Measurement,
  author       = {H. Poddar and J. Zhang and X. Liu and M. Shafi},
  title        = {Measurement Campaigns, Datasets, and Curve Fitting Officially
                  Used by {3GPP} in the Release 19 for Channel Modeling in
                  {TR~38.901} for {7--24~GHz}},
  howpublished = {arXiv:2603.25927},
  year         = {2026},
}

@article{Thomas2025THzSurvey,
  author  = {Sidharth Thomas and Jaskirat Singh Virdi and Aydin Babakhani and Ian P. Roberts},
  title   = {A Survey on Advancements in {THz} Technology for {6G}: Systems, Circuits, Antennas, and Experiments},
  journal = {IEEE Open J. Commun. Soc.},
  volume  = {6},
  pages   = {1998--2016},
  year    = {2025},
  doi     = {10.1109/OJCOMS.2025.3549710}
}

@techreport{TR38901,
  author      = {{3GPP}},
  title       = {{TR} 38.901: Study on Channel Model for Frequencies from 0.5 to 100 {GHz}},
  institution = {3rd Generation Partnership Project (3GPP)},
  number      = {TR 38.901, V19.3.0},
  month       = mar,
  year        = {2026},
  url         = {https://www.3gpp.org/ftp/Specs/archive/38\_series/38.901/38901-j30.zip}
}

@article{sun2016investigation,
  author  = {Shu Sun and Theodore S. Rappaport and Timothy A. Thomas and Amitava Ghosh and Huan C. Nguyen and Istvan Z. Kovacs and Ignacio Rodriguez and Ozge Koymen and Andrzej Partyka},
  title   = {Investigation of Prediction Accuracy, Sensitivity, and Parameter Stability of Large-Scale Propagation Path Loss Models for {5G} Wireless Communications},
  journal = {IEEE Trans. Veh. Technol.},
  volume  = {65},
  number  = {5},
  pages   = {2843--2860},
  year    = {2016},
  doi     = {10.1109/TVT.2016.2543139}
}

@article{xing2021millimeter,
  author  = {Yunchou Xing and Theodore S. Rappaport},
  title   = {Millimeter Wave and Terahertz Urban Microcell Propagation Measurements and Models},
  journal = {IEEE Commun. Lett.},
  volume  = {25},
  number  = {12},
  pages   = {3755--3759},
  year    = {2021},
  doi     = {10.1109/LCOMM.2021.3117900}
}

@article{shakya2024radio,
  author  = {Dipankar Shakya and Shihao Ju and Ojas Kanhere and Hitesh Poddar and Yunchou Xing and Theodore S. Rappaport},
  title   = {Radio Propagation Measurements and Statistical Channel Models for Outdoor Urban Microcells in Open Squares and Streets at 142, 73, and 28 {GHz}},
  journal = {IEEE Trans. Antennas Propag.},
  volume  = {72},
  number  = {4},
  pages   = {3580--3595},
  year    = {2024},
  doi     = {10.1109/TAP.2024.3366581}
}

@article{maccartney2019millimeter,
  author  = {George R. MacCartney and Theodore S. Rappaport},
  title   = {Millimeter-Wave Base Station Diversity for {5G} Coordinated Multipoint ({CoMP}) Applications},
  journal = {IEEE Trans. Wireless Commun.},
  volume  = {18},
  number  = {7},
  pages   = {3395--3410},
  year    = {2019},
  doi     = {10.1109/TWC.2019.2913414}
}

@article{itwillwork,
  author  = {Theodore S. Rappaport and Shu Sun and Rimma Mayzus and Hang Zhao and Yaniv Azar and Kevin Wang and George N. Wong and Jocelyn K. Schulz and Mathew Samimi and Felix Gutierrez},
  title   = {Millimeter Wave Mobile Communications for {5G} Cellular: It Will Work!},
  journal = {IEEE Access},
  volume  = {1},
  pages   = {335--349},
  year    = {2013},
  doi     = {10.1109/ACCESS.2013.2260813}
}

@inproceedings{xing2021umi,
  author    = {Yunchou Xing and Theodore S. Rappaport},
  title     = {Propagation Measurements and Path Loss Models for sub-{THz} in Urban Microcells},
  booktitle = {Proc. IEEE Int. Conf. Commun. (ICC)},
  year      = {2021},
  pages     = {1--6},
  doi       = {10.1109/ICC42927.2021.9500385}
}

@book{montgomery2012linear,
  author    = {Douglas C. Montgomery and Elizabeth A. Peck and G. Geoffrey Vining},
  title     = {Introduction to Linear Regression Analysis},
  edition   = {5},
  publisher = {John Wiley \& Sons},
  address   = {Hoboken, NJ, USA},
  year      = {2012},
  isbn      = {978-0-470-54281-1}
}

@inproceedings{sun2016vtc_pl,
  author    = {Shu Sun and Theodore S. Rappaport and Sundeep Rangan and Timothy A. Thomas and Amitava Ghosh and Istvan Z. Kovacs and Ignacio Rodriguez and Ozge Koymen and Andrzej Partyka and Jan Jarvelainen},
  title     = {Propagation Path Loss Models for {5G} Urban Micro- and Macro-Cellular Scenarios},
  booktitle = {Proc. IEEE 83rd Veh. Technol. Conf. (VTC2016-Spring)},
  year      = {2016},
  doi       = {10.1109/VTCSpring.2016.7504435}
}

@article{rappaport2015wideband,
  author  = {Theodore S. Rappaport and George R. {MacCartney Jr.} and Mathew K. Samimi and Shu Sun},
  title   = {Wideband Millimeter-Wave Propagation Measurements and Channel Models for Future Wireless Communication System Design},
  journal = {IEEE Trans. Commun.},
  volume  = {63},
  number  = {9},
  pages   = {3029--3056},
  year    = {2015},
  doi     = {10.1109/TCOMM.2015.2434384}
}

@article{rappaport2019above100ghz,
  author  = {Theodore S. Rappaport and Yunchou Xing and Ojas Kanhere and Shihao Ju and Arjuna Madanayake and Soumyajit Mandal and Ahmed Alkhateeb and Georgios C. Trichopoulos},
  title   = {Wireless Communications and Applications Above 100 {GHz}: Opportunities and Challenges for {6G} and Beyond},
  journal = {IEEE Access},
  volume  = {7},
  pages   = {78729--78757},
  year    = {2019},
  doi     = {10.1109/ACCESS.2019.2921522}
}

@inproceedings{ju2021spatial142ghz,
  author    = {Shihao Ju and Theodore S. Rappaport},
  title     = {Sub-Terahertz Spatial Statistical {MIMO} Channel Model for Urban Microcells at 142 {GHz}},
  booktitle = {Proc. IEEE Global Commun. Conf. (GLOBECOM)},
  year      = {2021},
  doi       = {10.1109/GLOBECOM46510.2021.9685929}
}

@article{rappaport2017overview,
  author  = {Theodore S. Rappaport and Yunchou Xing and George R. {MacCartney Jr.} and Andreas F. Molisch and Evangelos Mellios and Jianhua Zhang},
  title   = {Overview of Millimeter Wave Communications for Fifth-Generation ({5G}) Wireless Networks---With a Focus on Propagation Models},
  journal = {IEEE Trans. Antennas Propag.},
  volume  = {65},
  number  = {12},
  pages   = {6213--6230},
  year    = {2017},
  doi     = {10.1109/TAP.2017.2734243}
}

@inproceedings{sun2014multibeam,
  author    = {Shu Sun and George R. MacCartney and Mathew K. Samimi and Shuai Nie and Theodore S. Rappaport},
  title     = {Millimeter Wave Multi-Beam Antenna Combining for {5G} Cellular Link Improvement in {New York City}},
  booktitle = {Proc. IEEE Int. Conf. Commun. (ICC)},
  year      = {2014},
  pages     = {5468--5473},
  doi       = {10.1109/ICC.2014.6884203}
}

@inproceedings{nie2013indoor,
  author    = {Shuai Nie and George R. MacCartney and Shu Sun and Theodore S. Rappaport},
  title     = {{72 GHz} Millimeter Wave Indoor Measurements for Wireless and Backhaul Communications},
  booktitle = {Proc. IEEE Int. Symp. Pers., Indoor Mobile Radio Commun. (PIMRC)},
  year      = {2013},
  pages     = {2429--2433},
  doi       = {10.1109/PIMRC.2013.6666553}
}

@article{panjwani1996coverage,
  author  = {Mahesh A. Panjwani and A. Lynn Abbott and Theodore S. Rappaport},
  title   = {Interactive Computation of Coverage Regions for Wireless Communication in Multifloored Indoor Environments},
  journal = {IEEE J. Sel. Areas Commun.},
  volume  = {14},
  number  = {3},
  pages   = {420--430},
  year    = {1996},
  doi     = {10.1109/49.490227}
}

@misc{rappaport2004buildingDatabase,
  author       = {Theodore S. Rappaport and Roger Skidmore},
  title        = {Method and System for a Building Database Manipulator},
  howpublished = {U.S. Patent 6,721,769},
  year         = {2004}
}

@misc{rappaport2004receptionSurfaces,
  author       = {Theodore S. Rappaport and Roger Skidmore},
  title        = {System and Method for Ray Tracing Using Reception Surfaces},
  howpublished = {U.S. Patent Application 10/830,445},
  year         = {2004}
}

@inproceedings{samimi2015multifreq,
  author    = {Mathew K. Samimi and Theodore S. Rappaport},
  title     = {Statistical Channel Model with Multi-Frequency and Arbitrary Antenna Beamwidth for Millimeter-Wave Outdoor Communications},
  booktitle = {Proc. IEEE Global Commun. Conf. Workshops (GC Wkshps)},
  year      = {2015},
  pages     = {1--7},
  doi       = {10.1109/GLOCOMW.2015.7414164}
}

@inproceedings{shakya2025milcom,
  author    = {Dipankar Shakya and Naveed A. Abbasi and Mingjun Ying and Isha Jariwala and Jason J. Qin and Ishaan S. Gupte and Bridget Meier and Guanyue Qian and Daniel Abraham and Theodore S. Rappaport and Andreas F. Molisch},
  title     = {Standardized Machine-Readable Point-Data Format for Consolidating Wireless Propagation Across Environments, Frequencies, and Institutions},
  booktitle = {Proc. IEEE Military Commun. Conf. (MILCOM)},
  year      = {2025},
  pages     = {232--237},
  doi       = {10.1109/MILCOM64451.2025.11309979}
}

@article{ying2026nyurayCalibration,
  author  = {Mingjun Ying and Dipankar Shakya and Peijie Ma and Guanyue Qian and Theodore S. Rappaport},
  title   = {Site-Specific Location Calibration and Validation of Ray-Tracing Simulator {NYURay} at Upper Mid-Band Frequencies},
  journal = {npj Wireless Technol.},
  volume  = {2},
  number  = {1},
  pages   = {8},
  year    = {2026},
  doi     = {10.1038/s44459-025-00014-x}
}

@inproceedings{ying2025locationOptimization,
  author    = {Mingjun Ying and Peijie Ma and Dipankar Shakya and Theodore S. Rappaport},
  title     = {Multi-Stage Location Optimization Through Power Delay Profile Alignment Using Site-Specific Wireless Ray Tracing},
  booktitle = {Proc. IEEE Global Commun. Conf. (GLOBECOM)},
  year      = {2025},
  pages     = {5169--5174},
  doi       = {10.1109/GLOBECOM59602.2025.11432293}
}

@book{rappaport2002wireless,
  author    = {Theodore S. Rappaport},
  title     = {Wireless Communications: Principles and Practice},
  edition   = {2},
  publisher = {Prentice Hall},
  address   = {Upper Saddle River, NJ, USA},
  year      = {2002},
  isbn      = {0-13-042232-0}
}

@book{rappaport2015mmwavebook,
  author    = {Theodore S. Rappaport and Robert W. {Heath Jr.} and Robert C. Daniels and James N. Murdock},
  title     = {Millimeter Wave Wireless Communications},
  publisher = {Prentice Hall},
  address   = {Upper Saddle River, NJ, USA},
  year      = {2015},
  isbn      = {978-0-13-217228-8}
}

\end{document}